\pdfoutput=1

\documentclass[aps,prd,preprint,amsmath,amssymb,nofootinbib]{revtex4-1}

\def\ket#1{\langle #1 \rangle}

\DeclareMathOperator{\B}{B}
\DeclareMathOperator{\Conf}{Conf}
\DeclareMathOperator{\Gr}{Gr}
\DeclareMathOperator{\Li}{Li}

\begin{document}

\title{A Cluster Bootstrap for Two-Loop MHV Amplitudes}

\author{John Golden}
\author{Marcus Spradlin}

\affiliation{Department of Physics, Brown University\\
Box 1843, Providence, RI 02912-1843, USA}

\begin{abstract}

We apply a bootstrap procedure to two-loop MHV amplitudes
in planar $\mathcal{N}=4$ super-Yang-Mills theory.
We argue that the mathematically most complicated part
(the $\Lambda^2 \B_2$ coproduct
component) of the $n$-particle amplitude is uniquely determined by
a simple cluster algebra property together with
a few physical constraints (dihedral symmetry,
analytic structure, supersymmetry, and well-defined collinear limits).
We present a concise, closed-form expression
which manifests these properties for all $n$.

\end{abstract}

\maketitle

The scattering amplitudes of planar $\mathcal{N}=4$ supersymmetric Yang-Mills (SYM) theory~\cite{Brink:1976bc} comprise a collection of functions with remarkable mathematical properties, tightly restricted by the physical constraints they must satisfy. Indeed the mathematical and physical properties of these amplitudes are, collectively, so restrictive that one marvels that the functions can even exist at all.  The program of using known (or supposed) general properties of amplitudes to assemble concrete new results, which can then be verified by applying consistency checks, is generally known as the amplitude bootstrap.  One of the ultimate goals of this program is to formulate a concise list of simple physical and mathematical constraints which might uniquely determine the precise functional form of all amplitudes in SYM theory.

The simplest incarnation of the bootstrap program applies to those $L$-loop $n$-particle N${}^k$MHV amplitudes which belong to the class of generalized polylogarithm functions~\cite{G02}.
All amplitudes with $L<2$ or $n<10$ or $k<3$ are
believed~\cite{ArkaniHamed:2012nw} to belong to this class. Tools for dealing with the classes of functions which might appear in more general amplitudes are currently lacking, but there is every reason to suspect that a bootstrap program will forge ahead once the appropriate techniques are developed.

The essential tool for dealing with amplitudes of the generalized polylogarithm type is the symbol map (see~\cite{Goncharov:2009}).  The symbol of an amplitude of weight $w$ ($=2L$)
is an element of the $w$-fold tensor product of the multiplicative group generated by certain rational functions on the kinematic configuration space~\cite{Golden:2013xva} $\Conf_n(\mathbb{P}^3) = \Gr(4,n)/(\mathbb{C}^*)^{n-1}$ for $n$-particle scattering in SYM theory.
A fundamental working hypothesis of the bootstrap
is that the set of rational functions allowed to appear in the symbol of any amplitude (i.e., the ``symbol alphabet'') is the set of $\mathcal{A}$-coordinates on the $\Gr(4,n)$ cluster algebra.
Starting with the special collection of functions (or symbols) of this type, one bootstraps an amplitude
by applying constraints and comparing with independent data from the literature
(for example, from the amplitude OPE expansion~\cite{Basso:2013vsa,Sever:2011da,Gaiotto:2011dt,Alday:2010ku},
or from multi-Regge limits~\cite{Bartels:2008ce,Bartels:2008sc,Fadin:2011we,Lipatov:2012gk}),
until one arrives at a unique putative result.
The cluster $\mathcal{A}$-coordinate hypothesis is supported by all explicit results
for amplitudes available in the literature to date, including two-loop
MHV for all $n$~\cite{CaronHuot:2011ky}, two-loop NMHV for $n=6,7$~\cite{Dixon:2011nj,CaronHuot:2011kk},
three-loop MHV and NMHV
for
$n=6$~\cite{Dixon:2014iba,Dixon:2013eka}, and four-loop MHV for $n=6$~\cite{Dixon:2014voa}.

Some recent investigations~\cite{Golden:2014xqa,Golden:2013xva,Golden:2013lha,Golden:2014xqf} have revealed that
the connection between the cluster structure~\cite{FG03b} on $\Conf_n(\mathbb{P}^3)$ and the mathematical structure of amplitudes in SYM theory runs much deeper than merely specifying the appropriate symbol alphabet.
We use the term ``cluster bootstrap'' to emphasize that our focus will be on understanding the implications of these ``more clustery'' properties, and in particular on how to harness their power via bootstrap.
In this paper our attention is focused specifically
on the planar $n$-particle two-loop MHV amplitudes $R_n^{(2)}$.
We argue that these amplitudes are completely determined (modulo
classical polylogarithm functions $\Li_k$) by a straightforward cluster
property together with a few simple physical constraints.
Consideration of these constraints leads us to the concise explicit
formula~(\ref{eq:npt}) which specifies $R_n^{(2)}$ modulo $\Li_k$'s.

\section{Cluster Coordinates and Coproducts}

Let us begin by recalling a few relevant facts about the $\Gr(4,n)$ Grassmannian cluster algebra.
Physicists seeking additional background may find~\cite{Golden:2013xva} useful.
Cluster $\mathcal{X}$-coordinates are a preferred set of cross-ratios
(dual conformally invariant~\cite{Drummond:2008vq} ratios of
products of homogeneous polynomials in the $\mathcal{A}$-coordinates) on the kinematic
domain $\Conf_n(\mathbb{P}^3)$.  A ``cluster'' is a collection $\{x_i \}$ of $3(n-5)$ such coordinates
with the property that the Poisson bracket matrix $B_{ij} = \{ \log x_i, \log x_j \}$
has integer entries and maximal (if $n$ is odd) or nearly maximal (if $n$ is even) rank.
Cluster $\mathcal{X}$-coordinates may be systematically constructed via a process called mutation, and
a given $\mathcal{X}$-coordinate may appear in one or more clusters.
For $n=6, 7$ iterated application of mutations close on a finite set of clusters (14 and 833, respectively)
containing a finite number of distinct $\mathcal{X}$-coordinates (15, 385).

For $n>7$ one can mutate indefinitely to produce an infinite number of $\mathcal{A}$- and $\mathcal{X}$-coordinates, but
this poses no conceptual obstacle to the bootstrap program since only finitely many can appear
in any individual generalized polylogarithm function (i.e., at any finite loop order).
For example, the symbol of $R_n^{(2)}$ contains $\frac{n}{2}(3 n^2-30n + 77)$ $\mathcal{A}$-coordinates.  These can easily be enumerated by inspecting the all-$n$ result of~\cite{CaronHuot:2011ky}: in the notation of that paper, there are $n(n-6)$ symbol letters of the form $\langle 1(23)(n{-}1\,n)(i\,i{+}1)\rangle$ (plus all cyclic partners), $\frac{n}{2}(n-6)(n-7)$ of the form $\langle 12\,\overline{i} \cap \overline{j} \rangle$ (plus cyclic), and $\frac{n}{2}(n-5)(n-6)$ of the form $\langle 1(n2)(i\,i{+}1)(j\,j{+}1)\rangle$ (plus cyclic). Finally, there are of course the simple Pl\"ucker coordinates $\langle ijkl \rangle$, which number $\binom{n}{4}$; however it is evident from~\cite{CaronHuot:2011ky} that the only ones which appear in the two-loop MHV amplitudes are those in which at least one pair among $ij$, $jk$, $kl$ or $li$ are cyclically adjacent (for example, $\ket{1357}$ does not appear for $n>7$), so we must subtract $\frac{n}{24}(n-5)(n-6)(n-7)$ from $\binom{n}{4}$.  Adding up all of these types we find a total of $\frac{n}{2}(3 n^2-30n + 77)$ symbol letters.

To determine whether a dual conformal cross-ratio $R$ formed from these letters is an $\mathcal{X}$-coordinate, we apply a simple heuristic, originally described in~\cite{Golden:2013xva,Golden:2013lha}: 
$R$ is an $\mathcal{X}$-coordinate
if $1+R$ can also be expressed as a ratio of products of letters and
if $R$ is positive everywhere inside the positive domain (i.e., the domain in which $\ket{ijkl}> 0$ for all
$i<j<k<l$).  We know of no example where this heuristic fails.

Although the
symbol of $R_n^{(2)}$ is known for all $n$~\cite{CaronHuot:2011ky},
explicit analytic results for the function $R_n^{(2)}$ are available only for
$n=6,7$~\cite{Goncharov:2010jf,Golden:2014xqf}.
While obtaining more general fully analytic results is certainly a worthwhile
goal, in order to cut to the core of the mathematical structure of these
functions it is natural to focus on the coproduct (or, more properly,
the cobracket~\cite{Golden:2013xva})
\begin{equation}
\delta(R_n^{(2)}) \in \Lambda^2 \B_2 \bigoplus \B_3 \otimes \mathbb{C}^*.
\end{equation}
We remind the reader that the Bloch group $\B_n$ is generated
by elements denoted $\{x\}_n$.  Concretely, $\{x\}_n$ denotes the
equivalence class of weight-$n$ functions, modulo products, containing
$-\Li_n(-x)$.
Any generalized polylogarithm of weight 4 is determined, modulo products
of functions of lower weight, by the two coproduct components shown above.
Moreover the $\Lambda^2 \B_2$
component captures the ``most nontrivial'' part of a function and
determines the $\B_3 \otimes \mathbb{C}^*$ component
modulo terms involving the quadrilogarithm function $\Li_4$~\cite{G91a,G91b}.

In principle one could compute the coproduct $\delta(R_n^{(2)})$
directly from the known symbol of $R_n^{(2)}$.
However, the representation of the symbol given in~\cite{CaronHuot:2011ky}
does not have the appropriate cluster structure manifest, making such
a calculation infeasible.
Instead we put aside our knowledge of the symbol for a moment
while we bootstrap our way towards an explicit formula for
$\delta(R_n^{(2)})\rvert_{\Lambda^2 \B_2}$, shown in eq.~(\ref{eq:npt}).
While this formula is strictly speaking conjectural, being based on
some presumed cluster algebra structure of the amplitude, we have
checked it by explicit comparison to the results of~\cite{CaronHuot:2011ky}
through $n=13$.

\section{Elements of the Cluster Bootstrap}

The structure of MHV scattering amplitudes is heavily constrained at the level of the symbol, a fact which Dixon and collaborators have
exploited to great effect (see~\cite{Dixon:2014xca} for a recent review on the $n=6$ bootstrap).
We wish to adopt a similar approach at the level of
the coproduct, specifically the $\Lambda^2 \B_2$ component.  To this end we start
by formulating a hypothesis for what kinds of variables $x,y$ the cluster bootstrap should
allow to
appear in $\{x\}_2 \wedge \{y\}_2$.
It has been observed ``experimentally'', for
small values of $n$~\cite{Golden:2013xva}, that the
two-loop MHV amplitudes have the very special feature that their
$\Lambda^2 \B_2$ coproduct components are always expressible in terms of
linear combinations of terms $\{v\}_2 \wedge \{z\}_2$ where
\begin{enumerate}
\item{each $v$ is drawn from a set of cluster $\mathcal{X}$-coordinates
specially adapted to the analytic structure of the amplitude (related to
what is called the ``first-entry condition''),}
\item{each $z$ is drawn from a set of cluster $\mathcal{X}$-coordinates
specially adapted to the supersymmetry properties of MHV amplitudes
(related to what is called the ``last-entry condition''),}
\item{and the two variables $v, z$ appearing in each term always
belong to at least one cluster in common (this means, in particular,
that $\{ \log v, \log z\} \in  \{ -1, 0, +1\}$ with respect
to the natural Poisson bracket on the kinematic manifold
$\Conf_n(\mathbb{P}^3)$).
}
\end{enumerate}
Let us explain these points in a little more detail.

First we recall that the analytic structure of
color-ordered scattering amplitudes is highly constrained:  they may only have branch points
on the boundary of the Euclidean region at points where some sum of cyclically adjacent momenta
becomes null.
Requiring that amplitudes have only physical singularities implies that the first entries of the
symbol of any amplitude
must be drawn from the set of cross-ratios given by
\begin{equation}
u_{ij} = \frac{\ket{i\,i{+}1\,j{+}1\,j{+}2} \ket{i{+}1\,i{+}2\,j\,j{+}1}}
{\ket{i\,i{+}1\,j\,j{+}1} \ket{i{+}1\,i{+}2\,j{+}1\,j{+}2}}.
\end{equation}
Unfortunately, none of the $u_{ij}$ are cluster $\mathcal{X}$-coordinates.  Instead we consider
the closely related quantities
\begin{equation}
v_{ijk} = \frac{1}{\prod_{a=j}^{k-1} u_{ia}}-1
=-\frac{\langle i{+}1 (i\,i{+}2)(j\,j{+}1)(k\,k{+}1) \rangle}{\langle i\,i{+}1\,k\,k{+}1\rangle \langle i{+}1\,i{+}2\,j\,j{+}1\rangle},
\end{equation}
where 
\begin{equation}
\ket{a(bc)(de)(fg)} \equiv \ket{abde}\ket{acfg} - \ket{abfg} \ket{acde}.
\end{equation} $v_{ijk}$ is a $\mathcal{X}$-coordinates as long as $i<j<k$ (mod $n$). We can phrase the familiar first-entry
condition in terms of these unfamiliar variables by saying that only the quantities $1+v_{ijk}$ are allowed
in the first entry of the symbol of any function with physical branch cuts.

Secondly we recall the MHV last-entry condition~\cite{CaronHuot:2011kk}, which states that the last entry of the symbol of any MHV amplitude must, as a consequence of extended supersymmetry, be drawn from the set of Plu\"cker coordinates of the form $\ket{\overline{i}\,j} \equiv \ket{i{-}1\,i\,i{+}1\,j}$.
We therefore might like to include ratios built purely out of these objects in our ansatz. Unfortunately, no $\mathcal{X}$-coordinates of this type exist. Instead we consider the cross-ratios
\begin{equation}
z^+_{ijk} =\frac{\ket{i\,i{+}1\,\overline{j}\cap\overline{k}}}{\ket{i\,\overline{k}}\ket{i{+}1\,\overline{j}}},
\qquad
z^-_{ijk}=\frac{\langle i\,i{+}1\,j\,k\rangle \langle \overline{i}\,i{+}2\rangle}{\langle \overline{i}\,k \rangle \langle \overline{i{+}1} \,j\rangle},
\end{equation}
where 
\begin{equation}
	\ket{ab\overline{c} \cap \overline{d}} \equiv \ket{a\overline{c}}\ket{b\overline{d}} -
\ket{b\overline{c}}\ket{a\overline{d}}.
\end{equation} The $z^{\pm}_{ijk}$ are all
cluster $\mathcal{X}$-coordinates for $\Gr(4,n)$ as long as $i<j<k$ (mod $n$), and as suggested by the notation, $z^\pm_{ijk}$ are parity conjugates of each other (see~\cite{Golden:2013xva} for a discussion of
parity on $\Conf_n(\mathbb{P}^3)$).
Despite appearances these are in fact intimately tied to the last-entry condition since
\begin{equation}
1+z^+_{ijk} = \frac{\ket{i\,\overline{j}}\ket{i{+}1\,\overline{k}}}{\ket{i\,\overline{k}}\ket{i{+}1\,\overline{j}}},
\qquad
1+z^-_{ijk} = \frac{\ket{\overline{i}\,j}\ket{\overline{i{+}1}\,k}}{\ket{\overline{i}\,k}\ket{\overline{i{+}1}\,j}}.
\end{equation}
It is useful to define certain boundary cases of the above cross-ratios with overlapping indices:
\begin{equation}
v_{ij} = v_{i\,j\,j{+}1}, \qquad z_{ij} = z^-_{i\,j\,j{+}1},
\end{equation}
where parity takes $z_{ij}\to z_{ji}$.
Similar to what was done in the previous paragraph, we may express the familiar last-entry
condition in terms of these unfamiliar variables by saying that only the quantities $1+z_{ijk}^\pm$
are allowed in the final entry of the symbol of any MHV amplitude.

Third, it is worth commenting on how the Poisson bracket $\left\{ \log x,
\log y \right\}$ between two $\mathcal{X}$-coordinates
may be computed in practice.
If one could enumerate all possible clusters (and the
Poisson bracket matrix in each cluster), by repeated application
of the mutation algorithm starting
with the initial cluster reviewed in~\cite{Golden:2014xqf}, then one could scan that list to determine
whether or not a given pair $x, y$ appears together inside any cluster (and, if so, then one could read off their Poisson bracket).
For the infinite algebras we encounter when $n>7$ it is obviously not
feasible to enumerate all clusters.  An alternative approach would be to
express $x$ and $y$ as algebraic functions of the $\mathcal{X}$-coordinates
$(u_1,u_2,\ldots)$ in the initial cluster and then to compute
\begin{equation}
\{ \log x, \log y \} = \sum_{i,j} \frac{\partial \log x}{\partial \log u_i} \frac{\partial \log y}{\partial \log u_j} \{ \log u_i, \log u_j \}.
\end{equation}
If for a given pair $x,y$ the right-hand side comes out to be $0$ or $\pm 1$,
then it is guaranteed that there exists a cluster containing both $x$ and
$y$, even if it would be computationally infeasible to find a specific
path of mutations connecting that cluster to the initial cluster.
However, we have found that
the simplest way to compute $\{\log x, \log y\}$ for general
$x,y$ is to use the fact that the Poisson bracket on $\Gr(k,n)$ is
induced from the easily computible Sklyanin bracket on SL${}_n$, as
described for
example in~\cite{1057.53064}\footnote{We are very grateful to C.~Vergu for
pointing out this method to us.}.

The collection of all
$\frac{1}{2} n(n-5)^2$
of the $v$'s and $n(n-5)^2$ of the $z$'s constitutes what we call the $\{v,z\}$ basis.
Noting that $\{1+x\}_2 = - \{x\}_2$, the discussion in the previous two paragraphs suggests that
it is natural to seek a representation for $\delta(R_n^{(2)})\rvert_{\Lambda^2 \B_2}$
as a linear combination of objects of the form $\{v\}_2 \wedge \{z\}_2$ which capture, at the level of
the coproduct, the spirit of both the first- and last-entry constraints satisfied by the symbol.
Of course, the $\wedge$-product obscures any precise notion of first or last entries for the coproduct,
so our argument for restricting to
$\{v\}_2\wedge \{z\}_2$ is meant to be suggestive rather than rigorous.  The suitability of this ansatz is
justified {\it a posteriori} because it leads to a successful bootstrap.

Based on these considerations, as well as explicit calculations at small $n$, we are motivated
to hypothesize that
properties 1--3 listed above are true for general $n$, so we adopt these as
core elements of the cluster bootstrap for $R_n^{(2)}$.
In addition we impose that $R_n^{(2)}$ should be
\begin{enumerate}
\item[4.]{invariant under the dihedral group acting on the $n$ particle
labels, as well as under parity, and}
\item[5.]{well-defined under collinear limits.}
\end{enumerate}
We have found that these five simple physical and mathematical conditions
uniquely fix
$\delta(R^{(2)}_n)\rvert_{\Lambda^2 \B_2}$
(up to a single overall multiplicative constant common to all $n$)
to take the value shown explicitly in eq.~(\ref{eq:npt}).
Let us emphasize that in step 5 it is not actually necessary to know
the $n{-}1$-particle result in order to construct the answer
for $n$ particles; it is sufficient merely to make an appropriate ansatz
for the latter and impose only that the $n \parallel n{-}1$ collinear limit is well-defined.
This determines both the $n$- and $n{-}1$-particle results
at the same time, and in particular ties together their overall normalizations.

\section{Applying the bootstrap}

Let us explain in some detail how the procedure works beginning with $n=7$.
In this case there are are 14 $v$'s and 28 $z$'s, so we start with the ansatz that
$\delta(R^{(2)}_7)\rvert_{\Lambda^2 \B_2}$ should be a linear combination of
the $14\times 28=392$ possible $\{v\}_2\wedge\{z\}_2$'s. Only 70 of these pairs have Poisson brackets in the set $\{-1,0,+1\}$ (i.e., appear together in a cluster),
and after imposing dihedral and parity symmetries we are left with the three-parameter ansatz
\begin{equation}
\Big(
   c_1
   \left\{v_{14}\right\}_2\wedge
   \left\{z_{14}\right\}_2+c_2
   \left\{v_{14}\right\}_2\wedge \left\{z_{15}\right\}_2
+c_3 \left\{v_{146}\right\}_2\wedge
   \left\{z^{-}_{624}\right\}_2
\Big)
+ \text{ dihedral + conjugate}.
\end{equation}
We then take the collinear limit parameterized by
$Z_n \to Z_{n-1}+\alpha (Z_{n-2} + \beta Z_{1})+\gamma Z_{2}$
with $\gamma \to 0$, then $\alpha \to 0$, leaving $\beta$ free. This leads to
\begin{multline}
c_2 \Big((\left\{v_{25}v_{2\beta}/(1+v_{2\beta})\right\}_2
   +\left\{v_{25}/(1+v_{3\beta})\right\}_2)\wedge
   \left\{z_{36}\right\}_2\\ +\left\{v_{14}\right\}_2\wedge
   \left(\left\{z_{14}\right\}_2+\left\{z_{25}\right\}_2\right) +\left\{v_{36}\right\}_2\wedge
   \left(\left\{z_{14}\right\}_2+\left\{z_{36}\right\}_2\right)\Big)\\
   +(c_1-c_3)
 \Big(\left\{v_{2\beta}\right\}_2\wedge \left\{z_{25}\right\}_2-\left\{v_{3\beta}\right\}_2\wedge
   \left\{z_{36}\right\}_2\Big)+\text{conjugate},
\end{multline}
where $v_{i \beta} = \beta \ket{1\,i\,i{+}1\,6}/\ket{i\,i{+}1\,5\,6}$ and ``+ conjugate'' applies to the entire expression. For the collinear limit to be well-defined, i.e., independent of $\beta$ (which
specifies the relative length of the collinear momenta 6 and 7),
we require $c_1 = c_3$ and $c_2 = 0$. We have therefore determined that $\delta(R^{(2)}_{6})|_{\Lambda^2 \B_2} =0$ and also reproduce
the result~\cite{Golden:2013xva} that
\begin{equation}\label{eq:7pt}
 \delta(R^{(2)}_{7})|_{\Lambda^2 \B_2}=
   \left\{v_{14}\right\}_2\wedge
   \left\{z_{14}\right\}_2+ \left\{v_{146}\right\}_2\wedge
   \left\{z^{-}_{624}\right\}_2 + \text{ dihedral + conjugate},
 \end{equation}
up to an overall multiplicative constant.

The analogous ansatz for $n=8$ begins with $36 \times 72 = 2592$ terms of the form
$\{v\}_2 \wedge \{z\}_2$.
Restricting to pairs that appear together in a cluster reduces this to 400. After imposing the discrete symmetries only 15 free parameters remain, and requiring the $8 \parallel 7$ collinear limit to be well-defined
fixes all of them up to an overall normalization, which in turn may be fixed by matching
eq.~(\ref{eq:7pt}).
The bootstrap may be carried out through sufficiently large $n$ to motivate
the all-$n$ conjecture
\begin{widetext}
\begin{multline}\label{eq:npt}
\delta(R^{(2)}_{n})|_{\Lambda^2 \B_2} =\sum_{1< i<j< n}\Bigg( \{v_{1ij}\}_2\wedge\Big(-\{z_{ij}\}_2-\{z^-_{j2i}\}_2+\{z^-_{ij2}\}_2 +\{z^-_{j\,2\,i{+}1}\}_2-\{z^-_{i\,j{+}1\,2}\}_2\Big)\\ -\{v_{1i}\}_2 \wedge\Big(\{z^-_{j\,2\,i{+}1}\}_2 + \sum_{j<k\le 1} \{z_{jk}\}_2\Big) +\text{cyclic + conjugate}\Bigg).
\end{multline}
\end{widetext}
Here ``+ cyclic + conjugate'' applies to the both lines,
and we note that eq.~(\ref{eq:npt}) does satisfy the full dihedral symmetry
even though we have only chosen to manifest + cyclic.
The multiple sums contain some boundary terms which evaluate to $\{0\}_2$ or 
$\{\infty\}_2$; these are understood to be omitted.
We have explicitly checked (by comparing its iterated coproduct) that this expression is consistent
with the known symbol of $R_n^{(2)}$ through $n=13$.

\section{Discussion}

A striking and mysterious feature of eq.~(\ref{eq:npt}) is that all of the pairs $\{v,z\}$ appearing in the formula have Poisson bracket zero.
This feature is an output of the bootstrap; the input was much weaker, with the initial ansatz
also allowing pairs having Poisson bracket $\{ \log v, \log z\} = \pm 1$.
Geometrically, this means
that the $\Lambda^2 \B_2$ coproduct component wants to be expressed in terms of quadrilateral
(rather than pentagonal)
dimension-2 faces of the generalized Stasheff polytope associated to the $\Gr(4,n)$ cluster algebra,
as noted already in~\cite{Golden:2013xva}.

It would naturally be interesting to formulate a bootstrap for computing the $\B_3 \otimes \mathbb{C}^*$
coproduct components of $R_n^{(2)}$, which contain information about $\Li_4$ terms
that $\Lambda^2 \B_2$ does not know about.
Unfortunately we have found that
the $\{v,z\}$ basis provides an insufficient ansatz for $\B_3 \otimes \mathbb{C}^*$ already at
$n=7$.
Of course there is no obstacle to computing this coproduct component
on a case by case basis for small $n$ by starting with a larger collection of cluster $\mathcal{X}$-coordinates,
but deriving (or even guessing) an all-$n$ formula remains elusive.

It would also be interesting to extend eq.~(\ref{eq:npt}) to capture more or even all of
$R_n^{(2)}$, including terms involving products of $\Li_k$'s.
Interestingly we have found, using the standard symbol-level (anti-)symmetrization techniques outlined
in~\cite{Goncharov:2010jf}, that all of the $\Li_2 \Li_2$ terms in $R_n^{(2)}$ are captured by
the remarkable formula
\begin{equation}
   ``\text{eq.}~(\ref{eq:npt})" - \left(\sum_{1 \le i<j \le n} \Li_2(-v_{ij})\right)^2,
 \end{equation}
where the first term
refers to eq.~(\ref{eq:npt}) with the replacement $\{a\}_2 \wedge \{b\}_2 \to \Li_2(-a) \Li_2(-b)$.
This is a strong indication that
there is still more structure to discover in the two-loop $n$-particle MHV amplitudes.

Surely the most important and interesting
open question is whether a suitable ``cluster bootstrap'' can be formulated
for higher-loop MHV or non-MHV amplitudes.  The main obstacle is that so few explicit results
for such amplitudes are known, even just at the level of symbols, that we do not yet dare to speculate how
elements 1--3 of the cluster bootstrap ought to be generalized.

\section*{Acknowledgments}

We have benefitted from stimulating discussions and collaboration on closely related topics with
J.~Drummond, G.~Papathanasiou, M.~Paulos,
C.~Vergu, A.~Volovich, and especially A.~Goncharov.
This work was supported by the US Department of Energy under contract DE-SC0010010.

\bibliography{cluster_bootstrap}

\begin{thebibliography}{28}%
\makeatletter
\providecommand \@ifxundefined [1]{%
 \@ifx{#1\undefined}
}%
\providecommand \@ifnum [1]{%
 \ifnum #1\expandafter \@firstoftwo
 \else \expandafter \@secondoftwo
 \fi
}%
\providecommand \@ifx [1]{%
 \ifx #1\expandafter \@firstoftwo
 \else \expandafter \@secondoftwo
 \fi
}%
\providecommand \natexlab [1]{#1}%
\providecommand \enquote  [1]{``#1''}%
\providecommand \bibnamefont  [1]{#1}%
\providecommand \bibfnamefont [1]{#1}%
\providecommand \citenamefont [1]{#1}%
\providecommand \href@noop [0]{\@secondoftwo}%
\providecommand \href [0]{\begingroup \@sanitize@url \@href}%
\providecommand \@href[1]{\@@startlink{#1}\@@href}%
\providecommand \@@href[1]{\endgroup#1\@@endlink}%
\providecommand \@sanitize@url [0]{\catcode `\\12\catcode `\$12\catcode
  `\&12\catcode `\#12\catcode `\^12\catcode `\_12\catcode `\%12\relax}%
\providecommand \@@startlink[1]{}%
\providecommand \@@endlink[0]{}%
\providecommand \url  [0]{\begingroup\@sanitize@url \@url }%
\providecommand \@url [1]{\endgroup\@href {#1}{\urlprefix }}%
\providecommand \urlprefix  [0]{URL }%
\providecommand \Eprint [0]{\href }%
\providecommand \doibase [0]{http://dx.doi.org/}%
\providecommand \selectlanguage [0]{\@gobble}%
\providecommand \bibinfo  [0]{\@secondoftwo}%
\providecommand \bibfield  [0]{\@secondoftwo}%
\providecommand \translation [1]{[#1]}%
\providecommand \BibitemOpen [0]{}%
\providecommand \bibitemStop [0]{}%
\providecommand \bibitemNoStop [0]{.\EOS\space}%
\providecommand \EOS [0]{\spacefactor3000\relax}%
\providecommand \BibitemShut  [1]{\csname bibitem#1\endcsname}%
\let\auto@bib@innerbib\@empty
\bibitem [{\citenamefont {Brink}\ \emph {et~al.}(1977)\citenamefont {Brink},
  \citenamefont {Schwarz},\ and\ \citenamefont {Scherk}}]{Brink:1976bc}%
  \BibitemOpen
  \bibfield  {author} {\bibinfo {author} {\bibfnamefont {L.}~\bibnamefont
  {Brink}}, \bibinfo {author} {\bibfnamefont {J.~H.}\ \bibnamefont {Schwarz}},
  \ and\ \bibinfo {author} {\bibfnamefont {J.}~\bibnamefont {Scherk}},\ }\href
  {\doibase 10.1016/0550-3213(77)90328-5} {\bibfield  {journal} {\bibinfo
  {journal} {Nucl.Phys.}\ }\textbf {\bibinfo {volume} {B121}},\ \bibinfo
  {pages} {77} (\bibinfo {year} {1977})}\BibitemShut {NoStop}%
\bibitem [{\citenamefont {Goncharov}(2005)}]{G02}%
  \BibitemOpen
  \bibfield  {author} {\bibinfo {author} {\bibfnamefont {A.}~\bibnamefont
  {Goncharov}},\ }\href {\doibase 10.1215/S0012-7094-04-12822-2} {\bibfield
  {journal} {\bibinfo  {journal} {Duke Math. J.}\ }\textbf {\bibinfo {volume}
  {128}},\ \bibinfo {pages} {209} (\bibinfo {year} {2005})},\ \Eprint
  {http://arxiv.org/abs/math/0208144} {math/0208144} \BibitemShut {NoStop}%
\bibitem [{\citenamefont {Arkani-Hamed}\ \emph {et~al.}(2012)\citenamefont
  {Arkani-Hamed}, \citenamefont {Bourjaily}, \citenamefont {Cachazo},
  \citenamefont {Goncharov}, \citenamefont {Postnikov},\ and\ \citenamefont
  {Trnka}}]{ArkaniHamed:2012nw}%
  \BibitemOpen
  \bibfield  {author} {\bibinfo {author} {\bibfnamefont {N.}~\bibnamefont
  {Arkani-Hamed}}, \bibinfo {author} {\bibfnamefont {J.~L.}\ \bibnamefont
  {Bourjaily}}, \bibinfo {author} {\bibfnamefont {F.}~\bibnamefont {Cachazo}},
  \bibinfo {author} {\bibfnamefont {A.~B.}\ \bibnamefont {Goncharov}}, \bibinfo
  {author} {\bibfnamefont {A.}~\bibnamefont {Postnikov}}, \ and\ \bibinfo
  {author} {\bibfnamefont {J.}~\bibnamefont {Trnka}},\ }\href@noop {} {\
  (\bibinfo {year} {2012})},\ \Eprint {http://arxiv.org/abs/1212.5605}
  {arXiv:1212.5605 [hep-th]} \BibitemShut {NoStop}%
\bibitem [{\citenamefont {Goncharov}()}]{Goncharov:2009}%
  \BibitemOpen
  \bibfield  {author} {\bibinfo {author} {\bibfnamefont {A.}~\bibnamefont
  {Goncharov}},\ }\href@noop {} {\ }\Eprint {http://arxiv.org/abs/0908.2238}
  {arXiv:0908.2238 [math.AG]} \BibitemShut {NoStop}%
\bibitem [{\citenamefont {Golden}\ \emph
  {et~al.}(2014{\natexlab{a}})\citenamefont {Golden}, \citenamefont
  {Goncharov}, \citenamefont {Spradlin}, \citenamefont {Vergu},\ and\
  \citenamefont {Volovich}}]{Golden:2013xva}%
  \BibitemOpen
  \bibfield  {author} {\bibinfo {author} {\bibfnamefont {J.}~\bibnamefont
  {Golden}}, \bibinfo {author} {\bibfnamefont {A.~B.}\ \bibnamefont
  {Goncharov}}, \bibinfo {author} {\bibfnamefont {M.}~\bibnamefont {Spradlin}},
  \bibinfo {author} {\bibfnamefont {C.}~\bibnamefont {Vergu}}, \ and\ \bibinfo
  {author} {\bibfnamefont {A.}~\bibnamefont {Volovich}},\ }\href {\doibase
  10.1007/JHEP01(2014)091} {\bibfield  {journal} {\bibinfo  {journal} {JHEP}\
  }\textbf {\bibinfo {volume} {1401}},\ \bibinfo {pages} {091} (\bibinfo {year}
  {2014}{\natexlab{a}})},\ \Eprint {http://arxiv.org/abs/1305.1617}
  {arXiv:1305.1617 [hep-th]} \BibitemShut {NoStop}%
\bibitem [{\citenamefont {Basso}\ \emph {et~al.}(2013)\citenamefont {Basso},
  \citenamefont {Sever},\ and\ \citenamefont {Vieira}}]{Basso:2013vsa}%
  \BibitemOpen
  \bibfield  {author} {\bibinfo {author} {\bibfnamefont {B.}~\bibnamefont
  {Basso}}, \bibinfo {author} {\bibfnamefont {A.}~\bibnamefont {Sever}}, \ and\
  \bibinfo {author} {\bibfnamefont {P.}~\bibnamefont {Vieira}},\ }\href
  {\doibase 10.1103/PhysRevLett.111.091602} {\bibfield  {journal} {\bibinfo
  {journal} {Phys.Rev.Lett.}\ }\textbf {\bibinfo {volume} {111}},\ \bibinfo
  {pages} {091602} (\bibinfo {year} {2013})},\ \Eprint
  {http://arxiv.org/abs/1303.1396} {arXiv:1303.1396 [hep-th]} \BibitemShut
  {NoStop}%
\bibitem [{\citenamefont {Sever}\ \emph {et~al.}(2011)\citenamefont {Sever},
  \citenamefont {Vieira},\ and\ \citenamefont {Wang}}]{Sever:2011da}%
  \BibitemOpen
  \bibfield  {author} {\bibinfo {author} {\bibfnamefont {A.}~\bibnamefont
  {Sever}}, \bibinfo {author} {\bibfnamefont {P.}~\bibnamefont {Vieira}}, \
  and\ \bibinfo {author} {\bibfnamefont {T.}~\bibnamefont {Wang}},\ }\href
  {\doibase 10.1007/JHEP11(2011)051} {\bibfield  {journal} {\bibinfo  {journal}
  {JHEP}\ }\textbf {\bibinfo {volume} {1111}},\ \bibinfo {pages} {051}
  (\bibinfo {year} {2011})},\ \Eprint {http://arxiv.org/abs/1108.1575}
  {arXiv:1108.1575 [hep-th]} \BibitemShut {NoStop}%
\bibitem [{\citenamefont {Gaiotto}\ \emph {et~al.}(2011)\citenamefont
  {Gaiotto}, \citenamefont {Maldacena}, \citenamefont {Sever},\ and\
  \citenamefont {Vieira}}]{Gaiotto:2011dt}%
  \BibitemOpen
  \bibfield  {author} {\bibinfo {author} {\bibfnamefont {D.}~\bibnamefont
  {Gaiotto}}, \bibinfo {author} {\bibfnamefont {J.}~\bibnamefont {Maldacena}},
  \bibinfo {author} {\bibfnamefont {A.}~\bibnamefont {Sever}}, \ and\ \bibinfo
  {author} {\bibfnamefont {P.}~\bibnamefont {Vieira}},\ }\href {\doibase
  10.1007/JHEP12(2011)011} {\bibfield  {journal} {\bibinfo  {journal} {JHEP}\
  }\textbf {\bibinfo {volume} {1112}},\ \bibinfo {pages} {011} (\bibinfo {year}
  {2011})},\ \Eprint {http://arxiv.org/abs/1102.0062} {arXiv:1102.0062
  [hep-th]} \BibitemShut {NoStop}%
\bibitem [{\citenamefont {Alday}\ \emph {et~al.}(2011)\citenamefont {Alday},
  \citenamefont {Gaiotto}, \citenamefont {Maldacena}, \citenamefont {Sever},\
  and\ \citenamefont {Vieira}}]{Alday:2010ku}%
  \BibitemOpen
  \bibfield  {author} {\bibinfo {author} {\bibfnamefont {L.~F.}\ \bibnamefont
  {Alday}}, \bibinfo {author} {\bibfnamefont {D.}~\bibnamefont {Gaiotto}},
  \bibinfo {author} {\bibfnamefont {J.}~\bibnamefont {Maldacena}}, \bibinfo
  {author} {\bibfnamefont {A.}~\bibnamefont {Sever}}, \ and\ \bibinfo {author}
  {\bibfnamefont {P.}~\bibnamefont {Vieira}},\ }\href {\doibase
  10.1007/JHEP04(2011)088} {\bibfield  {journal} {\bibinfo  {journal} {JHEP}\
  }\textbf {\bibinfo {volume} {1104}},\ \bibinfo {pages} {088} (\bibinfo {year}
  {2011})},\ \Eprint {http://arxiv.org/abs/1006.2788} {arXiv:1006.2788
  [hep-th]} \BibitemShut {NoStop}%
\bibitem [{\citenamefont {Bartels}\ \emph {et~al.}(2009)\citenamefont
  {Bartels}, \citenamefont {Lipatov},\ and\ \citenamefont
  {Sabio~Vera}}]{Bartels:2008ce}%
  \BibitemOpen
  \bibfield  {author} {\bibinfo {author} {\bibfnamefont {J.}~\bibnamefont
  {Bartels}}, \bibinfo {author} {\bibfnamefont {L.}~\bibnamefont {Lipatov}}, \
  and\ \bibinfo {author} {\bibfnamefont {A.}~\bibnamefont {Sabio~Vera}},\
  }\href {\doibase 10.1103/PhysRevD.80.045002} {\bibfield  {journal} {\bibinfo
  {journal} {Phys.Rev.}\ }\textbf {\bibinfo {volume} {D80}},\ \bibinfo {pages}
  {045002} (\bibinfo {year} {2009})},\ \Eprint {http://arxiv.org/abs/0802.2065}
  {arXiv:0802.2065 [hep-th]} \BibitemShut {NoStop}%
\bibitem [{\citenamefont {Bartels}\ \emph {et~al.}(2010)\citenamefont
  {Bartels}, \citenamefont {Lipatov},\ and\ \citenamefont
  {Sabio~Vera}}]{Bartels:2008sc}%
  \BibitemOpen
  \bibfield  {author} {\bibinfo {author} {\bibfnamefont {J.}~\bibnamefont
  {Bartels}}, \bibinfo {author} {\bibfnamefont {L.}~\bibnamefont {Lipatov}}, \
  and\ \bibinfo {author} {\bibfnamefont {A.}~\bibnamefont {Sabio~Vera}},\
  }\href {\doibase 10.1140/epjc/s10052-009-1218-5} {\bibfield  {journal}
  {\bibinfo  {journal} {Eur.Phys.J.}\ }\textbf {\bibinfo {volume} {C65}},\
  \bibinfo {pages} {587} (\bibinfo {year} {2010})},\ \Eprint
  {http://arxiv.org/abs/0807.0894} {arXiv:0807.0894 [hep-th]} \BibitemShut
  {NoStop}%
\bibitem [{\citenamefont {Fadin}\ and\ \citenamefont
  {Lipatov}(2012)}]{Fadin:2011we}%
  \BibitemOpen
  \bibfield  {author} {\bibinfo {author} {\bibfnamefont {V.}~\bibnamefont
  {Fadin}}\ and\ \bibinfo {author} {\bibfnamefont {L.}~\bibnamefont
  {Lipatov}},\ }\href {\doibase 10.1016/j.physletb.2011.11.048} {\bibfield
  {journal} {\bibinfo  {journal} {Phys.Lett.}\ }\textbf {\bibinfo {volume}
  {B706}},\ \bibinfo {pages} {470} (\bibinfo {year} {2012})},\ \Eprint
  {http://arxiv.org/abs/1111.0782} {arXiv:1111.0782 [hep-th]} \BibitemShut
  {NoStop}%
\bibitem [{\citenamefont {Lipatov}\ \emph {et~al.}(2013)\citenamefont
  {Lipatov}, \citenamefont {Prygarin},\ and\ \citenamefont
  {Schnitzer}}]{Lipatov:2012gk}%
  \BibitemOpen
  \bibfield  {author} {\bibinfo {author} {\bibfnamefont {L.}~\bibnamefont
  {Lipatov}}, \bibinfo {author} {\bibfnamefont {A.}~\bibnamefont {Prygarin}}, \
  and\ \bibinfo {author} {\bibfnamefont {H.~J.}\ \bibnamefont {Schnitzer}},\
  }\href {\doibase 10.1007/JHEP01(2013)068} {\bibfield  {journal} {\bibinfo
  {journal} {JHEP}\ }\textbf {\bibinfo {volume} {1301}},\ \bibinfo {pages}
  {068} (\bibinfo {year} {2013})},\ \Eprint {http://arxiv.org/abs/1205.0186}
  {arXiv:1205.0186 [hep-th]} \BibitemShut {NoStop}%
\bibitem [{\citenamefont {Caron-Huot}(2011)}]{CaronHuot:2011ky}%
  \BibitemOpen
  \bibfield  {author} {\bibinfo {author} {\bibfnamefont {S.}~\bibnamefont
  {Caron-Huot}},\ }\href {\doibase 10.1007/JHEP12(2011)066} {\bibfield
  {journal} {\bibinfo  {journal} {JHEP}\ }\textbf {\bibinfo {volume} {1112}},\
  \bibinfo {pages} {066} (\bibinfo {year} {2011})},\ \Eprint
  {http://arxiv.org/abs/1105.5606} {arXiv:1105.5606 [hep-th]} \BibitemShut
  {NoStop}%
\bibitem [{\citenamefont {Dixon}\ \emph {et~al.}(2012)\citenamefont {Dixon},
  \citenamefont {Drummond},\ and\ \citenamefont {Henn}}]{Dixon:2011nj}%
  \BibitemOpen
  \bibfield  {author} {\bibinfo {author} {\bibfnamefont {L.~J.}\ \bibnamefont
  {Dixon}}, \bibinfo {author} {\bibfnamefont {J.~M.}\ \bibnamefont {Drummond}},
  \ and\ \bibinfo {author} {\bibfnamefont {J.~M.}\ \bibnamefont {Henn}},\
  }\href {\doibase 10.1007/JHEP01(2012)024} {\bibfield  {journal} {\bibinfo
  {journal} {JHEP}\ }\textbf {\bibinfo {volume} {1201}},\ \bibinfo {pages}
  {024} (\bibinfo {year} {2012})},\ \Eprint {http://arxiv.org/abs/1111.1704}
  {arXiv:1111.1704 [hep-th]} \BibitemShut {NoStop}%
\bibitem [{\citenamefont {Caron-Huot}\ and\ \citenamefont
  {He}(2012)}]{CaronHuot:2011kk}%
  \BibitemOpen
  \bibfield  {author} {\bibinfo {author} {\bibfnamefont {S.}~\bibnamefont
  {Caron-Huot}}\ and\ \bibinfo {author} {\bibfnamefont {S.}~\bibnamefont
  {He}},\ }\href {\doibase 10.1007/JHEP07(2012)174} {\bibfield  {journal}
  {\bibinfo  {journal} {JHEP}\ }\textbf {\bibinfo {volume} {1207}},\ \bibinfo
  {pages} {174} (\bibinfo {year} {2012})},\ \Eprint
  {http://arxiv.org/abs/1112.1060} {arXiv:1112.1060 [hep-th]} \BibitemShut
  {NoStop}%
\bibitem [{\citenamefont {Dixon}\ and\ \citenamefont {von
  Hippel}(2014)}]{Dixon:2014iba}%
  \BibitemOpen
  \bibfield  {author} {\bibinfo {author} {\bibfnamefont {L.~J.}\ \bibnamefont
  {Dixon}}\ and\ \bibinfo {author} {\bibfnamefont {M.}~\bibnamefont {von
  Hippel}},\ }\href {\doibase 10.1007/JHEP10(2014)065} {\bibfield  {journal}
  {\bibinfo  {journal} {JHEP}\ }\textbf {\bibinfo {volume} {1410}},\ \bibinfo
  {pages} {65} (\bibinfo {year} {2014})},\ \Eprint
  {http://arxiv.org/abs/1408.1505} {arXiv:1408.1505 [hep-th]} \BibitemShut
  {NoStop}%
\bibitem [{\citenamefont {Dixon}\ \emph {et~al.}(2013)\citenamefont {Dixon},
  \citenamefont {Drummond}, \citenamefont {von Hippel},\ and\ \citenamefont
  {Pennington}}]{Dixon:2013eka}%
  \BibitemOpen
  \bibfield  {author} {\bibinfo {author} {\bibfnamefont {L.~J.}\ \bibnamefont
  {Dixon}}, \bibinfo {author} {\bibfnamefont {J.~M.}\ \bibnamefont {Drummond}},
  \bibinfo {author} {\bibfnamefont {M.}~\bibnamefont {von Hippel}}, \ and\
  \bibinfo {author} {\bibfnamefont {J.}~\bibnamefont {Pennington}},\ }\href
  {\doibase 10.1007/JHEP12(2013)049} {\bibfield  {journal} {\bibinfo  {journal}
  {JHEP}\ }\textbf {\bibinfo {volume} {1312}},\ \bibinfo {pages} {049}
  (\bibinfo {year} {2013})},\ \Eprint {http://arxiv.org/abs/1308.2276}
  {arXiv:1308.2276 [hep-th]} \BibitemShut {NoStop}%
\bibitem [{\citenamefont {Dixon}\ \emph
  {et~al.}(2014{\natexlab{a}})\citenamefont {Dixon}, \citenamefont {Drummond},
  \citenamefont {Duhr},\ and\ \citenamefont {Pennington}}]{Dixon:2014voa}%
  \BibitemOpen
  \bibfield  {author} {\bibinfo {author} {\bibfnamefont {L.~J.}\ \bibnamefont
  {Dixon}}, \bibinfo {author} {\bibfnamefont {J.~M.}\ \bibnamefont {Drummond}},
  \bibinfo {author} {\bibfnamefont {C.}~\bibnamefont {Duhr}}, \ and\ \bibinfo
  {author} {\bibfnamefont {J.}~\bibnamefont {Pennington}},\ }\href {\doibase
  10.1007/JHEP06(2014)116} {\bibfield  {journal} {\bibinfo  {journal} {JHEP}\
  }\textbf {\bibinfo {volume} {1406}},\ \bibinfo {pages} {116} (\bibinfo {year}
  {2014}{\natexlab{a}})},\ \Eprint {http://arxiv.org/abs/1402.3300}
  {arXiv:1402.3300 [hep-th]} \BibitemShut {NoStop}%
\bibitem [{\citenamefont {Golden}\ \emph
  {et~al.}(2014{\natexlab{b}})\citenamefont {Golden}, \citenamefont {Paulos},
  \citenamefont {Spradlin},\ and\ \citenamefont {Volovich}}]{Golden:2014xqa}%
  \BibitemOpen
  \bibfield  {author} {\bibinfo {author} {\bibfnamefont {J.}~\bibnamefont
  {Golden}}, \bibinfo {author} {\bibfnamefont {M.~F.}\ \bibnamefont {Paulos}},
  \bibinfo {author} {\bibfnamefont {M.}~\bibnamefont {Spradlin}}, \ and\
  \bibinfo {author} {\bibfnamefont {A.}~\bibnamefont {Volovich}},\ }\href@noop
  {} {\  (\bibinfo {year} {2014}{\natexlab{b}})},\ \Eprint
  {http://arxiv.org/abs/1401.6446} {arXiv:1401.6446 [hep-th]} \BibitemShut
  {NoStop}%
\bibitem [{\citenamefont {Golden}\ and\ \citenamefont
  {Spradlin}(2013)}]{Golden:2013lha}%
  \BibitemOpen
  \bibfield  {author} {\bibinfo {author} {\bibfnamefont {J.}~\bibnamefont
  {Golden}}\ and\ \bibinfo {author} {\bibfnamefont {M.}~\bibnamefont
  {Spradlin}},\ }\href {\doibase 10.1007/JHEP09(2013)111} {\bibfield  {journal}
  {\bibinfo  {journal} {JHEP}\ }\textbf {\bibinfo {volume} {1309}},\ \bibinfo
  {pages} {111} (\bibinfo {year} {2013})},\ \Eprint
  {http://arxiv.org/abs/1306.1833} {arXiv:1306.1833 [hep-th]} \BibitemShut
  {NoStop}%
\bibitem [{\citenamefont {Golden}\ and\ \citenamefont
  {Spradlin}(2014)}]{Golden:2014xqf}%
  \BibitemOpen
  \bibfield  {author} {\bibinfo {author} {\bibfnamefont {J.}~\bibnamefont
  {Golden}}\ and\ \bibinfo {author} {\bibfnamefont {M.}~\bibnamefont
  {Spradlin}},\ }\href {\doibase 10.1007/JHEP08(2014)154} {\bibfield  {journal}
  {\bibinfo  {journal} {JHEP}\ }\textbf {\bibinfo {volume} {1408}},\ \bibinfo
  {pages} {154} (\bibinfo {year} {2014})},\ \Eprint
  {http://arxiv.org/abs/1406.2055} {arXiv:1406.2055 [hep-th]} \BibitemShut
  {NoStop}%
\bibitem [{\citenamefont {Fock}\ and\ \citenamefont {Goncharov}(2009)}]{FG03b}%
  \BibitemOpen
  \bibfield  {author} {\bibinfo {author} {\bibfnamefont {V.~V.}\ \bibnamefont
  {Fock}}\ and\ \bibinfo {author} {\bibfnamefont {A.~B.}\ \bibnamefont
  {Goncharov}},\ }\href@noop {} {\bibfield  {journal} {\bibinfo  {journal}
  {Ann. Sci. \'Ec. Norm. Sup\'er. (4)}\ }\textbf {\bibinfo {volume} {42}},\
  \bibinfo {pages} {865} (\bibinfo {year} {2009})},\ \Eprint
  {http://arxiv.org/abs/math/0311245} {math/0311245} \BibitemShut {NoStop}%
\bibitem [{\citenamefont {Drummond}\ \emph {et~al.}(2010)\citenamefont
  {Drummond}, \citenamefont {Henn}, \citenamefont {Korchemsky},\ and\
  \citenamefont {Sokatchev}}]{Drummond:2008vq}%
  \BibitemOpen
  \bibfield  {author} {\bibinfo {author} {\bibfnamefont {J.}~\bibnamefont
  {Drummond}}, \bibinfo {author} {\bibfnamefont {J.}~\bibnamefont {Henn}},
  \bibinfo {author} {\bibfnamefont {G.}~\bibnamefont {Korchemsky}}, \ and\
  \bibinfo {author} {\bibfnamefont {E.}~\bibnamefont {Sokatchev}},\ }\href
  {\doibase 10.1016/j.nuclphysb.2009.11.022} {\bibfield  {journal} {\bibinfo
  {journal} {Nucl.Phys.}\ }\textbf {\bibinfo {volume} {B828}},\ \bibinfo
  {pages} {317} (\bibinfo {year} {2010})},\ \Eprint
  {http://arxiv.org/abs/0807.1095} {arXiv:0807.1095 [hep-th]} \BibitemShut
  {NoStop}%
\bibitem [{\citenamefont {Goncharov}\ \emph {et~al.}(2010)\citenamefont
  {Goncharov}, \citenamefont {Spradlin}, \citenamefont {Vergu},\ and\
  \citenamefont {Volovich}}]{Goncharov:2010jf}%
  \BibitemOpen
  \bibfield  {author} {\bibinfo {author} {\bibfnamefont {A.~B.}\ \bibnamefont
  {Goncharov}}, \bibinfo {author} {\bibfnamefont {M.}~\bibnamefont {Spradlin}},
  \bibinfo {author} {\bibfnamefont {C.}~\bibnamefont {Vergu}}, \ and\ \bibinfo
  {author} {\bibfnamefont {A.}~\bibnamefont {Volovich}},\ }\href {\doibase
  10.1103/PhysRevLett.105.151605} {\bibfield  {journal} {\bibinfo  {journal}
  {Phys.Rev.Lett.}\ }\textbf {\bibinfo {volume} {105}},\ \bibinfo {pages}
  {151605} (\bibinfo {year} {2010})},\ \Eprint {http://arxiv.org/abs/1006.5703}
  {arXiv:1006.5703 [hep-th]} \BibitemShut {NoStop}%
\bibitem [{\citenamefont {Goncharov}(1994)}]{G91a}%
  \BibitemOpen
  \bibfield  {author} {\bibinfo {author} {\bibfnamefont {A.~B.}\ \bibnamefont
  {Goncharov}},\ }in\ \href@noop {} {\emph {\bibinfo {booktitle} {Motives
  ({S}eattle, {WA}, 1991)}}},\ \bibinfo {series} {Proc. Sympos. Pure Math.},
  Vol.~\bibinfo {volume} {55}\ (\bibinfo  {publisher} {Amer. Math. Soc.},\
  \bibinfo {address} {Providence, RI},\ \bibinfo {year} {1994})\ pp.\ \bibinfo
  {pages} {43--96}\BibitemShut {NoStop}%
\bibitem [{\citenamefont {Goncharov}(1995)}]{G91b}%
  \BibitemOpen
  \bibfield  {author} {\bibinfo {author} {\bibfnamefont {A.}~\bibnamefont
  {Goncharov}},\ }\href {\doibase 10.1006/aima.1995.1045} {\bibfield  {journal}
  {\bibinfo  {journal} {Adv. Math.}\ }\textbf {\bibinfo {volume} {114}},\
  \bibinfo {pages} {197} (\bibinfo {year} {1995})}\BibitemShut {NoStop}%
\bibitem [{\citenamefont {Dixon}\ \emph
  {et~al.}(2014{\natexlab{b}})\citenamefont {Dixon}, \citenamefont {Drummond},
  \citenamefont {Duhr}, \citenamefont {von Hippel},\ and\ \citenamefont
  {Pennington}}]{Dixon:2014xca}%
  \BibitemOpen
  \bibfield  {author} {\bibinfo {author} {\bibfnamefont {L.~J.}\ \bibnamefont
  {Dixon}}, \bibinfo {author} {\bibfnamefont {J.~M.}\ \bibnamefont {Drummond}},
  \bibinfo {author} {\bibfnamefont {C.}~\bibnamefont {Duhr}}, \bibinfo {author}
  {\bibfnamefont {M.}~\bibnamefont {von Hippel}}, \ and\ \bibinfo {author}
  {\bibfnamefont {J.}~\bibnamefont {Pennington}},\ }\href@noop {} {\bibfield
  {journal} {\bibinfo  {journal} {PoS}\ }\textbf {\bibinfo {volume} {LL2014}},\
  \bibinfo {pages} {077} (\bibinfo {year} {2014}{\natexlab{b}})},\ \Eprint
  {http://arxiv.org/abs/1407.4724} {arXiv:1407.4724 [hep-th]} \BibitemShut
  {NoStop}%
\bibitem{1057.53064}
M.~Gekhtman, M.~Shapiro, and A.~Vainshtein, {\it {Cluster algebras and Poisson
  geometry}},  {\em Mosc. Math. J.} {\bf 3} (2003), no.~3 899--934.
\end{thebibliography}%

\end{document}